\documentclass[twocolumn,aip,jcp,showpacs,a4paper,10pt]{revtex4-1}

\usepackage[utf8]{inputenc}           %% definit le mode de codage du texte
\usepackage[T1]{fontenc}              %% permet d'utiliser les caracteres accentue
\usepackage[english]{babel}

\usepackage{amsmath,amssymb,amsthm}

\usepackage[scale=0.80]{geometry}   

\usepackage{siunitx}                  %% pour ecrire correctement les unites SI (mieux)
\usepackage{relsize}                  %% taille relative de caracteres
\usepackage{graphicx}                 %% permet d'importer des graphiques
\usepackage{booktabs}                 %% permet de tracer des separateurs plus jolis dans les tableaux
\usepackage{microtype}
\usepackage{kpfonts}
\newcommand{\mcal}{\mathcal}

\newcommand{\mrm} {\mathrm}

\newcommand{\tit} {\textit}
\newcommand{\tbf} {\textbf}
\newcommand{\trm} {\textrm}

\newcommand{\sig} {\sigma}

\newcommand{\de}  {\delta}
\newcommand{\De}  {\Delta}

\newcommand{\ve}  {\varepsilon}

\newcommand{\al}  {\alpha}

\newcommand{\be}  {\beta}

\newcommand{\ga}  {\gamma}

\newcommand{\la}  {\lambda}

\newcommand{\ka}  {\kappa}

\newcommand{\Acal}{\mathcal A}

\newcommand{\Ecal}{\mathcal E}

\newcommand{\Lcal}{\mathcal L}

\newcommand{\Ocal}{\mathrm{O}}

\newcommand{\Scal}{\mathcal S}

\newcommand{\Ebb}{\mathbb E}

\newcommand{\Nbb}{\mathbb N}

\newcommand{\p}{\prime}

\newcommand{\na}{\nabla}
\newcommand{\dd}{\mathrm d}

\newcommand{\tde}  {\tilde}

\newcommand{\vit}  {   v_{i,t}}
\newcommand{\pit}  {   p_{i,t}}

\newcommand{\qit}  {   q_{i,t}}

\newcommand{\veit} { \ve_{i,t}}

\newcommand{\pjt}  {   p_{j,t}}

\newcommand{\qjt}  {   q_{j,t}}

\newcommand{\vejt} { \ve_{j,t}}

\newcommand{\pin}  {   p_i^n}

\newcommand{\qin}  {   q_i^n}

\newcommand{\vein} { \ve_i^n}

\newcommand{\pjn}  {   p_j^n}

\newcommand{\vejn} { \ve_j^n}

\newcommand{\vij}  {   v_{ij}}

\newcommand{\rij}  {   r_{ij}}

\newcommand{\gaij} { \ga_{ij}}

\newcommand{\sigij}{\sig_{ij}}

\newcommand{\vijn}  {   v_{ij}^n}
\newcommand{\pijn}  {   p_{ij}^n}
\newcommand{\rijn}  {   r_{ij}^n}

\newcommand{\Gijn}  {   G_{ij}^n}

\newcommand{\gaijn} { \ga_{ij}^n}

\newcommand{\vijt}  {   v_{ij,t}}

\newcommand{\rijt}  {   r_{ij,t}}

\newcommand{\Wijt}  {   W_{ij,t}}

\newcommand{\gaijt} { \ga_{ij,t}}

\newcommand{\Ntraj}{N_{\rm traj}}
\newcommand{\Nit}  {N_{\rm iter}}
\newcommand{\Niter}{N_{\rm iter}}

\begin{document}

\title{New parallelizable schemes for integrating the Dissipative Particle Dynamics with Energy Conservation}

\author{Ahmed-Amine Homman}
\author{Jean-Bernard Maillet}
\affiliation{CEA, DAM, DIF, F-91297 Arpajon, France}
%\altaffiliation{Universit\'e Paris-Est, CERMICS (ENPC), INRIA, F-77455 Marne-la-Vall\'ee, France}

\author{Julien Roussel}
\author{Gabriel Stoltz}
\affiliation{Universit\'e Paris-Est, CERMICS (ENPC), INRIA, F-77455 Marne-la-Vall\'ee, France}

\begin{abstract}
This work presents new parallelizable numerical schemes for the integration of Dissipative Particle Dynamics with Energy conservation (DPDE). So far, no numerical scheme introduced in the literature is able to correctly preserve the energy over long times and give rise to small errors on average properties for moderately small timesteps, while being straightforwardly parallelizable. We present in this article two new methods, both straightforwardly parallelizable, allowing to correctly preserve the total energy of the system. We illustrate the accuracy and performance of these new schemes both on equilibrium and nonequilibrium parallel simulations.
\end{abstract}

\date{\today}

\maketitle

\newcommand{\SCALE}{1.00}
\newcommand{\SCALEXGR}{0.32}

\section{Introduction}

Molecular Dynamics (MD) nowadays easily allows the simulation of systems composed of several millions of atoms for molecular systems. This is however still insufficient when interesting phenomena occur on space and time scales spanning several orders of magnitude. One such example is the simulation of shock waves in molecular materials, where the resolution of atomic vibrations limits the timestep to fractions of picoseconds when covalent bonds are present. Coarse-grained models, which reduce the complexity of MD simulations with full atomistic details by representing some degrees of freedom in an average manner, are therefore of primary interest in this context, they allow to extend the spatial scale by reducing the number of degrees of freedom, and also the temporal scales by increasing the admissible timesteps since mesoparticles interact via smoother potentials.

The Dissipative Particle Dynamics (DPD)\cite{HoogerbruggeKoelman92} is a particle-based coarse-grained model in which atoms, molecules or even groups of molecules are represented by a single mesoscale DPD particle. The time evolution of the mesoscale particles is governed by a stochastic differential equation. Dissipative and random forces allow to take into account some effect of the missing degrees of freedom. DPD was put on a firm theoretical ground in Ref.~\onlinecite{EspanolWarren95}, and was applied to study the properties of various systems\cite{MoeendarbaryNG09,DischerOrtiz07,GoujonMalfreyt11}. 

However, DPD intrinsically is an equilibrium model, with a prescribed temperature. The temperature is fixed through a fluctuation/dissipation relation between the magnitudes of the friction and the fluctuation parameters. DPD cannot be used as such to study nonequilibrium systems. It should be replaced by a dynamics where the fluctuation-dissipation relation is based on variables which evolve in time. DPD with conserved energy (DPDE) is such a model\cite{AvalosMackie97,Espanol97}. 

In the DPDE framework, mesoparticles have an additional degree of freedom, namely an internal energy, which accounts for the energy of the missing degrees of freedom. The dynamics on the internal energies is constructed in order for the total energy of the system to remain constant. DPDE was for instance used to simulate shock and detonation waves\cite{Stoltz06,MSS07,MVDS11}.

However, the efficient numerical integration of DPDE still requires some effort. Numerous efficient schemes were developed for DPD\cite{GrootWarren97,PagonabarragaHagen98,BesoldVattulainen00,VattulainenKarttunen02} (see in particular Ref.~\onlinecite{LS15} for a careful comparison). However, their adaption to DPDE leads to numerical schemes for which errors on average properties may be large even for timesteps standardly used to integrate Hamiltonian dynamics\cite{LarentzosBrennan14}. This may therefore require the use of extremely small timesteps\cite{AbuNada10,AbuNada11}.

Among the currently known schemes for DPDE, one of them turns out to enjoy very nice properties in terms of energy conservation. It is a splitting scheme inspired from Shardlow's splitting scheme for DPD\cite{Shardlow03}, and adapted to the DPDE framework\cite{Stoltz06}. The resulting scheme, called SSA in this work (Shardlow Splitting Algorithm), is so far considered as the reference scheme for the numerical integration of DPDE\cite{LisalBrennan11}. The main issue with SSA is that particle pairs have to be updated sequentially. The parallelization of SSA is therefore not an easy task\cite{LarentzosBrennan14}. In addition, SSA is not threadable, and thus incompatible with the architectures of future supercomputers.

The aim of our work is to propose and test new schemes to integrate DPDE. We pay a particular attention to (i) the preservation of energy, (ii) the size of the errors on average properties arising from the use of finite timesteps, and (iii) the possibility to easily parallelize the method. More precisely, our aim is to construct easily parallelizable schemes for which the systematic errors (\emph{i.e.} the remaining errors in the limit of infinite sampling precision) are as small as possible. 

Let us also emphasize that the schemes we develop for DPDE may be of direct interest for other dynamics with similar structures such as the Smoothed Dissipative Particle Dynamics (SDPD)\cite{SDPD}. This stochastic dynamics also preserves the energy of the system. It can also be seen as the superposition of two elementary energy preserving dynamics, a conservative part and a fluctuation/dissipation part. It is therefore not a surprise that the numerical methods we develop here can be readily applied for large scale simulations of SDPD on parallel architectures\cite{FMS15}.

This article is organized as follows. We first briefly recall the DPDE model and its properties in Section~\ref{section:model_presentation}. We next present the numerical schemes we consider in Section~\ref{section:numerical_schemes}: some reference schemes in the literature are recalled in Section~\ref{subsection:existing_schemes}, while new schemes are proposed in Section~\ref{subsection:new_schemes}. We then perform sequential and parallel equilibrium simulations to evaluate the quality of the energy conservation and the errors on average properties as a function of the timestep (see Section~\ref{section:equilibrium_simulations}). Finally, in order to test the dynamical relevance of the newly introduced schemes, we present parallel simulations of shock waves in Section~\ref{section:shock_simulations}.

%------------------------------------------------------------------------
\section{Presentation of the model}
\label{section:model_presentation}

\subsection{The DPDE equations}
\label{subsection:DPDE}

We consider a system of $N$ identical particles in dimension $d$, with positions $q_i$ and momenta $p_i$. Their velocities are $v_i=\frac{p_i}m$, where $m$ is the mass of the particles (masses are taken to be identical to simplify the presentation; the extension to particles with different masses is straightforward). We rely on the DPDE model\cite{AvalosMackie97,Espanol97} although alternative deterministic models have also been proposed\cite{SH04}. In the DPDE framework, each particle represents a (group of) molecule(s), $q_i$ and $p_i$ being the position and momentum of the center of mass of this collection of atoms. These external degrees of freedom are coupled with some internal energies $\ve_i$ describing in an average manner the missing degrees of freedom lost in the coarse-graining process. Each pair of particles interacts through (i) conservative forces deriving from a potential energy function $U(q)$, (ii) friction forces proportional to the relative velocity between the two particles and (iii) random fluctuation forces. In this article, we consider a variant of the original DPDE model introduced in Ref.~\onlinecite{Stoltz06}, where friction and random forces are not projected along the line joining two particles. However, the numerical schemes presented in Section~\ref{section:numerical_schemes} can be straightforwardly modified for the original DPDE model.

The time evolution of the configuration $(q_i,p_i,\ve_i)$ of the $i$th DPDE particle is given by the following set of equations:
\begin{equation}
  \label{eq:DPDE}
  \left\{
  \begin{aligned}
    \dd  \qit & = \vit\,\dd t, \\ 
    \dd  \pit & = - \nabla_{q_i} U(q_t)\,\dd t - \sum_{j=1,j\neq i}^N \gaijt\chi(\rijt)\vijt\,\dd t \\
              & \ \ + \sum_{j=1,j\neq i}^N\sig\sqrt{\chi(\rijt)}\,\dd \Wijt, \\   
    \dd \veit  & = \frac 12\left[ \sum^N_{j=1,j\neq i} \left(\gaijt\vijt^2 - d\frac{\sig^2}m\right)\chi(\rijt)\,\dd t \right.\\
            &  \ \ \left. \phantom{\sum^N_{j=1,j\neq i}} -\sum^N_{j=1,j\neq i}\sig\sqrt{\chi(\rijt)}\vijt\cdot\dd \Wijt\right],
  \end{aligned}
  \right.
\end{equation}
where $(\Wijt)_{1 \leq i < j \leq N}$ are standard $d$-dimensional Brownian motions, and $W_{ij,t}=-W_{ji,t}$ for $i > j$. In the above equations, we have introduced the relative velocity $\vijt=v_{i,t}-v_{j,t}$ between particles $i$ and $j$ and their distance $\rijt=\|\qit-\qjt\|$. The function $\chi(r)$ is a cut-off function, limiting the range of the random and dissipative interactions. We consider here
\begin{equation}\label{eq:pot_mou}
 \chi(r) = \left\{\begin{aligned} 
 \left(1 - \frac r{r_{\rm cut}}\right)^2 & \quad \text{if $r\leq r_{\rm cut}$}, \\
 0                                     & \quad \text{otherwise}.
\end{aligned}\right.
\end{equation}
Note that, by construction, the total momentum $\sum_{i=1}^N p_i$ and the total energy 
\[
\Ecal(q,p,\ve) = U(q) + \sum_{i=1}^N \frac{p_i^2}m + \sum_{i=1}^N \ve_i
\]
are preserved. 

The parameters $\gaijt$ and $\sig$ respectively control the friction and fluctuation strengths. They are chosen in order to ensure the correct statistical behavior of the system. An important point is that the invariant measure should be consistent with the preservation of total energy and momentum. More precisely, $\gaijt$ and $\sig$ are chosen in order for measures
\begin{equation}\label{eq:DPDE_invariant_measure_general}
  f\Big(\Ecal(q,p,\ve)\Big)g\left(\sum_{i=1}^N p_i\right)\exp\Big(\Scal(\ve)\Big) \, \dd q \, \dd p \, \dd \ve
\end{equation} 
to be invariant under the dynamics~\eqref{eq:DPDE} for arbitrary functions~$f$ and~$g$. 
In this expression, 
\begin{displaymath}
\displaystyle \Scal(\ve) = \sum^N_{i=1} s(\ve_i)
\end{displaymath}
is the total internal entropy of the system, $s(\ve_i)$ being the internal entropy of the $i$th particle. The internal entropy function satisfies 
\[
s^{\p}(\ve) = \frac1{T(\ve)},
\]
where $T(\ve)$ is the internal temperature implicitly defined via the relation
\begin{equation}\label{eq:DPDE_internalEOS}
\displaystyle \ve = \int_0^{T(\ve)}C_v(\theta)\,\dd\theta.
\end{equation}
Here, $C_v(\theta)$ is the internal heat capacity at constant volume. Therefore, the internal heat capacity fully determines the internal energies and the internal entropies. The relation~\eqref{eq:DPDE_internalEOS} is the internal equation of state of the DPDE particle.

In order for measures of the form~\eqref{eq:DPDE_invariant_measure_general} to be invariant by the dynamics~\eqref{eq:DPDE}, the parameters $\gaijt$ and $\sig$ should satisfy the following fluctuation/dissipation relation\cite{Espanol97,AvalosMackie97}:
\begin{equation}\label{eq:DPDE_fd}
  \gamma_{ij}=\frac{\sig^2}4\left(\frac1{T_{i}}+\frac1{T_{j}}\right),
\end{equation}
where $T_i = T(\varepsilon_i)$ is the internal temperature of the $i$th particle. In order to highlight the similarity with the standard fluctuation/dissipation relation for Langevin dynamics (see for instance Section~2.2.3 in Ref.~\onlinecite{BookLelievreRousset}), we find it convenient to rewrite the definition of $\gamma_{ij}$ and $\sig$ as
\begin{equation}
  \label{eq:alternative_FDR}
  \gamma_{ij} = \ga\frac{\beta_{ij}}{\be_{\rm ref}},\qquad \sig=\sqrt{\frac{2\ga}{\be_{\rm ref}}},
\end{equation}
with 
\[
\beta_{ij}=\frac1{2k_{\rm B}}\left(\frac1{T_i}+\frac1{T_j}\right),
\]
$k_{\rm B}$ being Boltzmann's constant. This amounts to introducing a reference friction parameter~$\gamma$ and a reference temperature~$T_{\rm ref}$, with associated inverse energy $\be_{\rm ref}=\frac1{k_{\rm B}T_{\rm ref}}$.

\subsection{Thermodynamic properties}
\label{subsection:temperature_estimation}

In order to compute the average value of a physical observable $A$, we assume that the dynamics~\eqref{eq:DPDE} is ergodic for some invariant measure~$\mu$ of the form~\eqref{eq:DPDE_invariant_measure_general}. Note indeed that there are no mathematical results ensuring the ergodicity of the stochastic dynamics~\eqref{eq:DPDE} since the fluctuation part may be degenerate (contrarily to, say, Langevin dynamics). Ergodicity is even not known to hold for DPD, except in the very specific case of one-dimensional systems\cite{SY06}.

Since the total energy and the total momentum are preserved by the dynamics, the only reasonable assumption is that ergodicity holds with respect to the probability measure
\[
\begin{aligned}
& \mu_{E_0,P_0}(dq\,dp\,d\ve) \\
& \quad = Z^{-1}_{E_0,P_0} \delta_{ \{\Ecal(q,p,\ve)-E_0 \}} \delta_{ \left\{ \sum_{i=1}^N p_i - P_0 \right\}} \mrm{e}^{\Scal(\ve)},
\end{aligned}
\]
where $Z_{E_0,P_0}$ is a normalization constant. Under this assumption, average properties can be computed as
\begin{equation}\label{eq:ergodicity}
\left\langle A\right\rangle  = \int A \, \dd\mu_{E_0,P_0} = \lim_{t\to+\infty} \frac1t\int_0^t A(q_{\tau},p_{\tau},\ve_{\tau})\,\dd \tau,
\end{equation}
where $(q_{\tau},p_{\tau},\ve_{\tau})$ is the solution at time~$\tau$ of~\eqref{eq:DPDE} starting from a given configuration $(q_0,p_0,\ve_0)$. This initial configuration is such that $\Ecal(q_0,p_0,\ve_0) = E_0$ and $\sum_{i=1}^N p_{i,0} = P_0$.

The measure $\mu_{E_0,P_0}$ can be interpreted as some microcanonical measure. In order to obtain estimates of thermodynamic properties such as temperature, it is convenient to introduce a canonical equivalent of the measure~$\mu_{E_0,P_0}$. In fact, the measure $\mu_{E_0,0}$ (corresponding to a system with total momentum set to~0) should be equivalent in the limit of large systems to the canonical probability measure
\begin{equation}\label{eq:DPDE_ergodic_measure} 
    \dd\mu_{\be}(q,p,\ve) = Z_\be^{-1} \mrm{e}^{-\be \Ecal(q,p,\ve) + \Scal(\ve)}\, \dd q \, \dd p \, \dd \ve,
\end{equation}
where $Z_\be$ is a normalization constant. The parameter~$\be$ is chosen in order for the average energy under~\eqref{eq:DPDE_ergodic_measure} to be equal to~$E_0$:
\[
\int \Ecal(q,p,\ve) \, \dd\mu_{\be}(q,p,\ve) = E_0.
\]
This equivalence is very similar to the standard equivalence between microcanonical and canonical measures for systems described only by their positions and momenta.

We focus in the numerical tests we present on various estimators of the temperature, each one involving only one type of the three categories of degrees of freedom of the system: the positions $q_i$, the momenta $p_i$ and the internal energies $\ve_i$. The interest of these estimators is that they should all predict the same temperature (provided the system is sufficiently large, which, in our experience, is already the case for moderately large systems of several hundreds of particles). This therefore allows to assess the quality of the sampling for each group of degrees of freedom. 

The first estimator of the temperature is the standard kinetic temperature
\begin{equation}\label{eq:Tkin}
T_{\rm kin} = \frac 1{k_{\rm B}N_{\rm eff}} \left\langle \sum_{i=1}^N \frac{p_i^2}m\right\rangle_\beta,
\end{equation}
where $\langle \cdot \rangle_\beta$ refers to averages with respect to the measure~$\mu_\beta$ introduced in~\eqref{eq:DPDE_ergodic_measure}, and $N_{\rm eff}$ represents the effective number of external degrees of freedom of the system. It is \tit{a priori} equal to $dN$ but since we fix the total momentum to $P_0=0$, we reduce it to $N_{\rm eff}=d(N-1)$. This correction is anyway unimportant for sufficiently large systems.

The second estimator\cite{ButlerAyton98} depends only on the positions of the particles:
\begin{equation}\label{eq:Tcon}
\displaystyle T_{\rm pot} = \frac1{k_{\rm B}} \frac {\left\langle \|\nabla U\|^2\right\rangle_\beta } {\left\langle \De U\right\rangle_\beta }.
\end{equation}
Finally, the last estimator is determined by the internal energies: 
\begin{equation}\label{eq:Tint}
\displaystyle T_{\rm int} = \left(\left\langle \frac1{N}\sum_{i=1}^N \frac 1{T_i} \right\rangle_\beta\right)^{-1} .
\end{equation}
A simple computation shows that $T_{\rm pot} = T_{\rm int} = T_{\rm kin} = k_{\rm B}/\beta$ (the second equality is proved in Ref.~\onlinecite{MBAN99}). Let us emphasize that the internal temperature is estimated by a harmonic average rather than an arithmetic one. 

\subsection{Practical computation of average properties}
\label{subsection:computing_averages}

In order to estimate in practice average properties by the ergodic limit~\eqref{eq:ergodicity}, the dynamics needs to be numerically integrated. We therefore introduce a timestep $\De t > 0$, and denote by $(q^n,p^n,\ve^n)$ an approximation of the solution $(q_{n\De t},p_{n\De t},\ve_{n\De t})$ of~\eqref{eq:DPDE} obtained by iterating a numerical scheme. Average properties are then estimated by simulations over $\Niter$ timesteps as 
\begin{equation}\label{eq:obs_estimation}
    \displaystyle \left\langle A\right\rangle  \simeq \widehat{A}_{\De t,N_{\rm iter}} = \frac1{\Niter} \sum_{n=1}^{\Niter} A(q^n,p^n,\ve^n).
\end{equation}
Standard results from the numerical analysis of stochastic differential equations (see for instance Section~2.3.1 in Ref.~\onlinecite{BookLelievreRousset}) allow to quantify the errors produced by the approximation~\eqref{eq:obs_estimation} as follows:
\begin{equation}\label{eq:obs_estimation_theorique}
\widehat{A}_{\De t,N_{\rm iter}} = \left\langle A\right\rangle  + K_A\De t^\eta + \frac{R_{A,\De t}}{\sqrt{\Nit\De t}}.
\end{equation}
This equality highlights two sources of error in the computation of average properties: 
\begin{itemize}
\item a \emph{statistical error} $R_{A,\De t} / \sqrt{\Nit\De t}$, which arises from the finiteness of the number of iterations. Typically, a central limit theorem holds, so that $R_{A,\De t}$ follows some Gaussian distribution with a variance~$\sigma_{A,\De t}^2$ close to the variance of the time averages estimated with the underlying continuous dynamics. The important point is that the statistical error decreases as the inverse of the square-root of the physical simulation time. 
\item a \emph{systematic bias} $K_A\De t^\eta$, which is the residual error persisting in the limit of infinite sampling accuracy. This error arises from the use of finite timesteps. 
\end{itemize}
The statistical error only mildly depends on the numerical scheme at hand since the variance~$\sigma_{A,\De t}^2$ is at first order in~$\De t$ the variance of the continuous process\cite{LMS15}. As in Ref.~\onlinecite{LisalBrennan11}, our focus in this work is therefore rather on the systematic bias, which may be quite different for various numerical schemes. Our aim is to construct schemes for which the systematic bias is as small as possible. For a given maximal number of step $N_{\rm iter}$, this allows to integrate the dynamics with larger timesteps, and hence reduce faster the statistical error. 

%------------------------------------------------------------------------
\section{Integrating DPDE}
\label{section:numerical_schemes}

The numerical schemes we consider in this section are based on splitting techniques as presented in the DPD context\cite{SerranodeFabritiis06,Shardlow03}. The idea of splitting algorithms is to decompose the dynamics into several elementary subdynamics, which are sequentially integrated. Splitting schemes are interesting when the elementary dynamics are simple to numerically integrate. 

Although splitting schemes for stochastic dynamics are nowadays quite popular, some standard integration schemes for DPD do not fall into this class. For instance, some integrators were obtained by an ad-hoc modification of the Verlet scheme\cite{Verlet67} for the integration of Hamiltonian dynamics\cite{BesoldVattulainen00}. For completeness, we also consider in our numerical experiments the adaption of the Velocity-Verlet scheme to the DPDE context. This scheme is abbreviated SVV in the sequel for Stochastic Velocity-Verlet, and is made precise in Appendix~\ref{sec:SVV}.

Let us also immediately emphasize that a complication of DPDE compared to other stochastic dynamics such as DPD arises from the presence of two invariants of the dynamics, namely the total momentum and the total energy. Ideally, numerical schemes should preserve these invariants over very long times, at least approximately (as the energy is approximately preserved by appropriate discretizations of Hamiltonian dynamics).

We first briefly present in Section~\ref{subsection:existing_schemes} two splitting schemes already known by the community: a simple splitting scheme based on a Euler-Maruyama discretization of the dissipative part of~\eqref{eq:DPDE} (termed SEM in the sequel) and the adaption to the DPDE context of the splitting proposed by Shardlow in the DPD context\cite{Shardlow03,Stoltz06} (termed SSA in the sequel). The simple splitting scheme is very easy to parallelize but leads to very large errors in the estimated properties; while SSA is quite accurate but somewhat cumbersome to parallelize\cite{LarentzosBrennan14}. This motivates the introduction of two new straightforwardly parallelizable schemes in Sections~\ref{subsubsection:SER} and~\ref{subsubsection:Hybrid}.

\subsection{State of the Art}
    \label{subsection:existing_schemes}

The DPDE dynamics~\eqref{eq:DPDE} can be decomposed into a Hamiltonian evolution 
\begin{equation}\label{eq:newton}
\left\{\begin{aligned}
\dd q_{i,t} &  = \frac{p_{i,t}}m\,\dd t, \\
\dd p_{i,t} &  = -\nabla_{q_i} U(q_{i,t})\,\dd t,
\end{aligned}\right.
\end{equation}
and a fluctuation/dissipation part
\begin{equation}\label{eq:DPDE_dissipative}
\left\{\begin{aligned}
  \dd  \pit &=- \sum_{j=1,j\neq i}^N \gaijt       \chi(\rijt)\vij\,\dd t,                                            \\
            & \ \ \ + \sum_{j=1,j\neq i}^N \sig\sqrt{\chi(\rijt)}\,   \dd \Wijt,                                        \\   
  \dd \veit &= \frac 12\left[ \sum^N_{j=1,j\neq i} \left(\gaijt\vijt^2 - d\frac{\sig^2}m\right)\chi(\rijt)\,\dd t, \right.\\
            & \qquad \left. - \sum^N_{j=1,j\neq i}\sig\sqrt{\chi(\rijt)}\vijt\cdot\dd \Wijt\right].
\end{aligned}\right.
\end{equation}
All splitting schemes first integrate the Hamiltonian part with a standard Velocity-Verlet discretization\cite{Verlet67}. The difference therefore solely arises from the subsequent treatment of the fluctuation/dissipation part. 

\subsubsection{Splitting Explicit-Euler: SEM}
\label{subsubsection:SEM}

The SEM scheme integrates the fluctuation/dissipation with a simple Euler-Maruyama discretization:
\begin{equation}\label{eq:SEM_fd}
\left\{\begin{aligned}
p_i^{n+1} = \pin & - \sum^N_{j=1,j\neq i} \gaijn    \chi(\rijn)\vijn      \De t   \\
          & + \sum^N_{j=1,j\neq i} \sig\sqrt{\chi(\rijn)} \Gijn\sqrt{\De t}, \\
\ve_i^{n+1} = \vein & + \frac12 \sum^N_{j=1,j\neq i} \left(\gaijn(\vijn)^2-d\frac{\sig^2}{m}\right)\chi(\rijn)\De t  \\
           & - \frac12\sum^N_{j=1,j\neq i} \sig\sqrt{\chi(\rijn)}\vijn\cdot\Gijn\sqrt{\De t},
\end{aligned}\right.
\end{equation}
where (here and in the sequel) $(G^n_{ij})_{n\in\Nbb, 1 \leq i < j \leq N}$ are identically and independently distributed standard $d$-dimensional Gaussian random variables, and $\Gijn=-G^n_{ji}$. Note that no particular care is taken to make sure that the energy is preserved by the discretization of the fluctuation/dissipation part. Unsurprisingly, it turns out that the total energy drifts in time (see Section~\ref{subsection:energy_conservation}).

As SVV, SEM turns out not to be a very accurate scheme. It can be shown to be of weak order one. Under appropriate ergodicity conditions, the bias on average properties is therefore of order~$\De t$ ($\eta=1$ in \eqref{eq:obs_estimation_theorique}). We refer to Appendix~\ref{sec:scheme_consistency} for more precisions. 

The reason why we consider SVV and SEM in our numerical tests is that they can be efficiently parallelized in a straightforward manner, and are therefore appealing alternatives to SSA for large scale simulations on massively parallel architectures.

\subsubsection{Shardlow's Splitting Algorithm: SSA}
\label{subsubsection:SSA}

SSA, in opposition to SEM, further decomposes the fluctuation/dissipation dynamics~\eqref{eq:DPDE_dissipative} into elementary pairwise fluctuation/dissipation dynamics involving only two DPDE particles~$i$ and~$j$. Taking advantage of the energy and momentum conservations, these elementary fluctuation/dissipation equations can in fact be rewritten as follows:
\begin{equation}\label{eq:DPDE_ij}
\left\{\begin{aligned}
    \dd \pit  &\displaystyle= -\gaijt\vijt\chi(\rijt)\,\dd t + \sig\sqrt{\chi(\rijt)}\,\dd\Wijt, \\
    \dd \pjt  &\displaystyle= -\,\dd \pit,\\
    \dd \veit &\displaystyle= -\frac12 \,\dd\left( \frac{\pit^2}{2m} + \frac{\pjt^2}{2m} \right), \\
    \dd \vejt &\displaystyle= \,\dd\veit.
\end{aligned}\right.
\end{equation}
As noted in Ref.~\onlinecite{SerranodeFabritiis06} for DPD, the dynamics on the momenta can be analytically integrated when $\gaij$ is fixed (\emph{i.e} for fixed $\ve_i$, $\ve_j$), instead of using a modified BBK algorithm\cite{BruengerBrooks84} as done in Ref.~\onlinecite{Shardlow03}. For the extension of SSA to DPDE we consider in this work, we rely on the strategy introduced in Ref.~\onlinecite{Stoltz06}, where momenta are updated first (here with an analytical integration at fixed~$\gaij$), and internal energies are then updated to ensure the overall energy conservation. Introducing 
\[
a_{ij,\De t} = -\frac{2\gaij}{m}\chi(\rij)\De t,
\]
the scheme we use to integrate the elementary pairwise fluctuation/dissipation~\eqref{eq:DPDE_ij} reads
\begin{equation}\label{eq:SSA_ij}
\left\{\begin{aligned}
    p_i^{n+1} = \pin & + \frac12\Bigg[ \left(\mrm{e}^{-a^n_{ij,\De t}}-1\right) \pijn   \\
                   & \quad + \left. \sig\sqrt{ \frac{m\left(1-\mrm{e}^{-2a^n_{ij,\De t}}\right)}{\gaijn} }\Gijn \right], \\
    p_j^{n+1} = \pjn & - \frac12\Bigg[ \left(\mrm{e}^{-a^n_{ij,\De t}}-1\right) \pijn   \\
                   & \quad + \left. \sig\sqrt{ \frac{m\left(1-\mrm{e}^{-2a^n_{ij,\De t}}\right)}{\gaijn} }\Gijn \right], \\
    \ve_i^{n+1} = \vein & - \frac12\De K^n_{ij}, \\
    \ve_j^{n+1} = \vejn & - \frac12\De K^n_{ij},
\end{aligned}\right.
\end{equation}
where we introduced the kinetic energy variation 
\[
\De K_{ij}^n = \frac1{2m}\left[(p_i^{n+1})^2 + (p_j^{n+1})^2 - (\pin)^2 - (\pjn)^2 \right]
\] 
in order to update the internal energies. Particle couples which are sufficiently close to interact are then sequentially updated with the scheme~\eqref{eq:SSA_ij}. It is precisely this sequentiality which prevents a simple parallelization of the scheme. As a side remark, note that when linear scaling techniques are used (such as decomposing the system into cells of size $r_{\rm cut}$), the order in which the particles are updated may change from one iteration to the other.

\subsection{New Numerical schemes for integrating DPDE}
\label{subsection:new_schemes}

The two new parallelizable schemes we propose rely on the splitting between Hamiltonian and fluctuation/dissipation parts mentioned at the beginning of Section~\ref{subsection:existing_schemes}, but with new strategies to discretize the fluctuation/dissipation part:
\begin{itemize}
\item The first scheme we present in Section~\ref{subsubsection:SER} is called ``Splitting with Energy Reinjection'' (termed SER in the sequel). Its discretization of the dissipative part~\eqref{eq:DPDE_fd} is similar to SEM but uses a global symmetric reinjection of the kinetic energy variation into the internal energies, instead of directly discretizing the dynamics of the internal energies. This allows to automatically preserve the total energy during the fluctuation/dissipation part.
\item The second scheme is a mix between SSA and SER, and is therefore termed ``Hybrid'' in the sequel (see Section~\ref{subsubsection:Hybrid}). As parallel simulations are performed using a spatial repartition of the simulation box between the processors, the bottleneck for the parallelization of SSA arises from particles located in different domains \cite{LarentzosBrennan14}. Therefore, the idea of the Hybrid scheme is to integrate the elementary fluctuation/dissipations interactions involving particles located on the same processor by a pass of the SSA algorithm, while the remaining interactions are taken care of by a SER discretization. 
\end{itemize}

\subsubsection{Splitting with Energy Reinjection: SER}
\label{subsubsection:SER}

The SER integration of the fluctuation/dissipation~\eqref{eq:DPDE_dissipative} is performed in two steps. First, momenta are integrated using a simple Euler-Maruyama discretization as
\begin{equation}\label{eq:SER_p}
    p^{n+1}_i = p^n_i + \de p^n_i, \quad \de p^n_i = \sum_{j=1,j\neq i}^N \de p^n_{ij},
\end{equation}
with $\de p^n_{ij} = -\ga_{ij}^n\chi(r^n_{ij})v^n_{ij}\De t + \sig\sqrt{\chi(r^n_{ij})}G^n_{ij}\sqrt{\De t}$. The internal energies $\ve_i$ are then updated in order to compensate for the energy variation during the update of the momenta. In order to implement this idea, we need to identify in the global kinetic energy variation of each particle the contribution of every single pairwise interaction, in order to redistribute the associated elementary energy variations. In fact, a simple computation shows that 
\[
\De K^n_i = \frac{(p^{n+1}_i)^2}{2m} - \frac{(p^n_i)^2}{2m} = \sum^N_{j=1,j\neq i} \De_j K^n_i,
\]
where 
\[
\De_j K^n_i = \de p^n_{ij} \cdot \left( v^n_i + \frac1{2m}\de p^n_i \right).
\]
represents the contribution of particle $j$ to the kinetic energy variation of particle $i$. The internal energies are then updated by reinjecting the elementary variations $\De_j K^n_i$ in a symmetric manner:
\begin{equation}\label{eq:SER_eint_discretization}
\begin{aligned}
\De \ve^n_i &= -\frac12 \sum_{j\neq i} \left( \De_iK^n_j + \De_jK^n_i \right)\\
           & = -\frac 12 \sum_{j\neq i} \de p^n_{ij} \cdot \left( v^n_{ij} + \frac1{2m}\left(\de p^n_i - \de p^n_j\right) \right).
\end{aligned}
\end{equation}
The discretization of equation \eqref{eq:DPDE_ij} can therefore be summarized as
\begin{equation}\label{eq:SER_fd}
\left\{\begin{aligned}
      p^{n+1}_i = \pin  &- \sum^N_{j=1,j\neq i} \gaijn\chi(\rijn)\vijn\De t\\
                & + \sum^N_{j=1,j\neq i} \sig\sqrt{\chi(\rijn)}\Gijn\sqrt{\De t}. \\
    \ve^{n+1}_i = \vein & - \sum^N_{j=1,j\neq i} \de p^n_{ij}\cdot\left( \vijn + \frac1{2m}\left(\de p_i^n - \de p_j^n\right) \right).
\end{aligned}\right.
\end{equation}
It can be shown that this numerical scheme is consistent. In fact, it has a weak order~1, so that the bias on average properties is of order~$\Delta t$. This can be checked by some straightforward but lengthy calculus, see Appendix~\ref{sec:consistency_SER}.

Let us emphasize that, by construction, SER automatically ensures the exact conservation of the total energy during the numerical integration of the fluctuation/dissipation part~\eqref{eq:DPDE_dissipative}. This very nice feature was only enjoyed by SSA among the previously known schemes to integrate DPDE. On the other hand, in opposition to SSA, SER does no rely on a further splitting of the fluctuation/dissipation into elementary pairwise interactions, and is therefore straightforward to parallelize. However, the presence of both $\de p^n_{ij}$ and $\de p^n_i$ in~\eqref{eq:SER_eint_discretization} requires two sweeps on the system, a first one to compute $\de p^n_{ij}$ and sum it into~$\de p^n_i$, and a second one to compute the products $\de p^n_{ij} \cdot \de p^n_i$. This seemingly requires to store all increments $\de p^n_{ij}$ (or at least the Gaussian increments $\Gijn$), which is somewhat prohibitive. This can however be avoided if the Gaussian increments $\Gijn$ can be exactly regenerated to the value they had on the first pass of the algorithm when the particle pairs are revisited. The SARU pseudo-random generator\cite{AfsharSchmid12} precisely allows such an easy recomputation since the Gaussian increments $\Gijn$ are completely determined by the integers $i,j$ and $n$.

\subsubsection{A scheme between SSA and SER: Hybrid}
\label{subsubsection:Hybrid}

The Hybrid scheme can be seen as a blending of SER and SSA, where the elementary fluctuation/dissipation interactions involving particles on the same processor are integrated by a pass of the SSA algorithm, while the remaining interactions are integrated with the SER scheme.

Let us describe more precisely this algorithm.
We denote by $\Phi^{\rm Hybrid}_{\De t}(q,p,\ve,G)$ the result of a Hybrid discretization of \eqref{eq:DPDE}, with timestep $\De t$, applied to the configuration $(q,p,\ve)$ of the system, i.e
\[ (q^{n+1},p^{n+1},\ve^{n+1}) =  \Phi^{\rm Hybrid}_{\De t}(q^n,p^n,\ve^n,G^n). \]
The variable $G = (G_{ij})_{i < j \leq N}$ refers to the Gaussian increments used for the discretization.
We also denote by $\Phi^{\trm{SSA},ij}_{\De t}(q,p,\ve,G)$ the result of a SSA discretization~\eqref{eq:SSA_ij} of the elementary dynamics~\eqref{eq:DPDE_ij} with timestep $\De t$.
Note that only the components $p_i,p_j,\ve_i,\ve_j$ of~$(q,p,\ve)$ are changed by $\Phi^{\trm{SSA},ij}_{\De t}(q,p,\ve,G)$.
Let us denote by $\Acal$ the set of particle couples where both particles are in the same processor, and $\Acal^c$ the remaining couples.
The elements of $\Acal$ are denoted by $(i_1,j_1)$, $(i_2,j_2)$, $\hdots$, $(i_l,j_l)$, where $l$ is the number of elements in $\Acal$.
Finally, we denote by $\Phi^{\trm{SER},\Acal^c}_{\De t}(q,p,\ve,G)$ the one-step iteration corresponding to a SER discretization of all the interactions involving the particle couples in $\Acal^C$.
The expression of $\Phi^{\rm Hybrid}_{\De t}(q,p,\ve,G)$ is obtained from $\Phi^{\trm{SSA},ij}_{\De t}(q,p,\ve,G)$ and $\Phi^{\trm{SER},\Acal^c}_{\De t}(q,p,\ve,G)$ as
\[
    \Phi^{\rm Hybrid}_{\De t}(q,p,\ve,G) = \Phi^{\trm{SER},\Acal^c}_{\De t} \circ \Phi^{\Acal}_{\De t} (q,p,\ve,G),
\]
with
\[
    \Phi^{\Acal}_{\De t}(q,p,\ve,G) = \Phi^{\trm{SSA},i_lj_l}_{\De t} \circ \hdots \circ \Phi^{\trm{SSA},i_1,j_1}_{\De t} (q,p,\ve,G),
\]
where $\circ$ denotes the mathematical composition.

The Hybrid scheme, in opposition to all the other schemes, does not depend only on the timestep but also depends on the space repartition of the simulation box between processors used for the simulations.
This dependence is studied in Section~\ref{subsection:parallelization_influence}.

%------------------------------------------------------------------------
\section{Equilibrium simulations}
\label{section:equilibrium_simulations}

In this section, we perform sequential and parallel equilibrium simulations using all the schemes presented in Section~\ref{section:numerical_schemes}.
However, the Hybrid scheme can only be tested in parallel simulations, otherwise it reduces to SSA. Therefore, no Hybrid scheme is used in the sequential simulations of Section~\ref{subsection:energy_conservation} and~\ref{subsubsection:sequential_simulations}.

\subsection{Description of the simulation}
\label{subsection:simulation_description}

We consider a system of $N$ particles, with pairwise conservative interactions described by a shifted, splined and truncated Lennard-Jones potential:
\begin{equation}\label{eq:Ep}
v(r) =\left\{\begin{array}{ll}
        \displaystyle 4\ve_{\rm LJ}\left( \left(\frac{r_{\rm LJ}}r\right)^{12} - \left(\frac{r_{\rm LJ}}r\right)^6 \right) &\\
\displaystyle \quad - \ve_{\rm sh} - f_{\rm sp}(r-r_{\rm cut})& \text{if $r \leq r_{\rm cut}$} \\
0 & \text{if $r \geq r_{\rm cut}$}.
\end{array}\right.
\end{equation}
The parameters $\ve_{\rm sh}$ and $f_{\rm sp}$ are chosen so that $v(r_{\rm cut})=0$ and $v^{\prime}(r_{\rm cut}) = 0$.
The total potential energy of the system reads
\begin{equation}\label{eq:def_potential_energy}
\displaystyle U(q) = \sum_{0\leq i<j \leq N} v(\rij).
\end{equation}
In all the simulations presented below, the heat capacity is supposed constant. Equation~\eqref{eq:DPDE_internalEOS} then simply reduces to $\ve_i = C_vT_i$.

All results are given in reduced units and, unless otherwise mentioned, the fluctuation strength is set to $\sigma = 4$. The physical dimension $d$ is set to~3 with the particle number set to respectively $N=343$, $N=2000$ and $N=144,000$ in Sections~\ref{subsection:energy_conservation},~\ref{subsubsection:sequential_simulations} and~\ref{subsubsection:parallel_simulations}.
We fix $C_v=50$ for the sequential simulations of Section~\ref{subsection:energy_conservation} and~\ref{subsubsection:sequential_simulations} and $C_v=10$ for the parallel simulations of Section~\ref{subsubsection:parallel_simulations}.

Initial conditions are obtained in Sections~\ref{subsection:energy_conservation} and~\ref{subsubsection:sequential_simulations} by melting a simple cubic crystal using a SSA scheme for $t=100$ with a timestep $\De t=0.001$. The initial condition for the parallel simulations of Section~\ref{subsubsection:parallel_simulations} is obtained by melting a cubic face centered crystal using a SER scheme for $t=45$ with a timestep $\De t=0.001$.
The density is $\rho=0.8095$ in all cases except in Section~\ref{subsection:energy_conservation} where $\rho=0.5787$.

\subsection{Energy drifts}
\label{subsection:energy_conservation}

The first task of an appropriate numerical scheme for DPDE is to preserve (at least approximately) the invariants of the dynamics, namely the energy and the total momentum. While the total momentum conservation is easily ensured, the energy conservation on the other hand is much more demanding. In fact, the numerical schemes we consider are obtained by a composition of a Verlet scheme, which approximately preserves the energy, and a discretization of the fluctuation/dissipation dynamics, which may or may not preserve the energy. The interaction between these two integrators may lead to drifts in the total energy, even when the integration of the fluctuation/dissipation part exactly preserves the energy, as already observed in Ref.~\onlinecite{LisalBrennan11}.

In order to precisely quantify the possible energy drifts and determine in particular the influence of the timestep $\De t$ on the drift rate, we compute the average evolution of the energy in time by performing $N_{\rm traj}=10^4$ trajectories of time $t_f = 10$ each. The initial conditions of each trajectory are sampled according to the canonical measure~$\mu_\beta$, obtained by sampling independently the internal energies according to the measure $Z_\varepsilon^{-1} \mathrm{e}^{-\beta \varepsilon + s(\varepsilon)}$, and the positions and momenta according to the canonical measure $Z_{q,p}^{-1} \mathrm{e}^{-\beta H(q,p)}$. We denote by $\left\langle A_t\right\rangle$ the value of an observable $A$ at time $t$, obtained by averaging over all the possible trajectories. In practice, $\left\langle A_{n\De t}\right\rangle$ is approximated by
\begin{equation}\label{eq:obs_time_estimation_traj}
  \widehat{A}^n_{\De t,\Ntraj} = \frac1\Ntraj\sum_{m=1}^{\Ntraj} A\left(x^{m,n}\right).
\end{equation}
where $x^{m,n}$ is the $n$-th configuration of the $m$-th trajectory.

\begin{figure}[h!]
\center
\includegraphics[scale=\SCALEXGR]{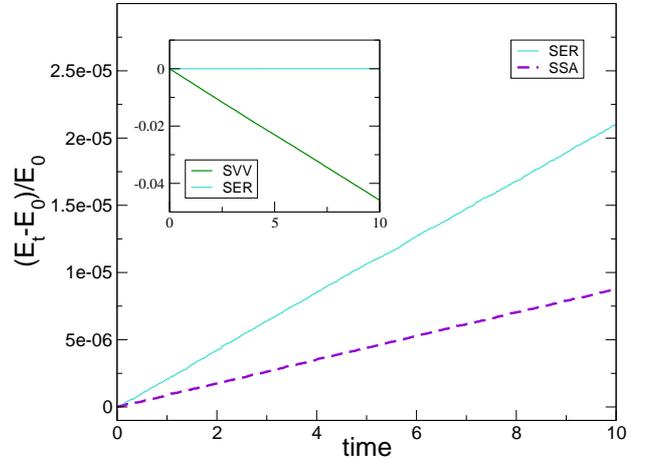}
\caption{\label{fig:E_t} Time-dependent energy drift averaged over trajectories, as a function of time when $\De t=0.006$.}
\end{figure}

Figure~\ref{fig:E_t} shows the behavior of the time-dependent energy drift averaged over trajectories, as a function of time, for various schemes and $\De t=0.006$.
Note that no scheme manages to keep the total energy constant. Linear drifts in time are observed in all cases, SVV being substantially worse than all the other schemes in terms of energy conservation. We study more precisely the drift rates in the remainder of this section, in order to determine the maximal timesteps which can be used in practice before average properties are corrupted.
     
\subsubsection{Quantification of the total energy drifts}
\label{subsubsection:energy_drifts}

\begin{figure}[h!]
\center
\includegraphics[scale=\SCALEXGR]{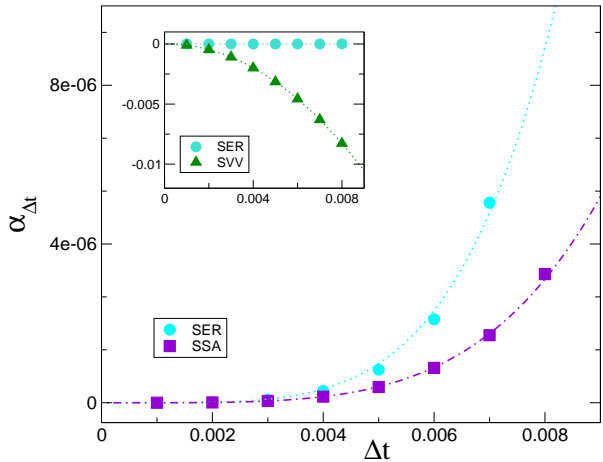}
\caption{\label{fig:dE_dt} Average energy drift rate $\al_{\De t}$, as a function of the timestep. A fit $\al_{\De t} = K\De t^\ka$ is superimposed to the data.}
\end{figure}

In order to decide whether the drift is acceptable, we quantify in this section the rate of variation of the energy as a function of the timestep $\De t$. Figure~\ref{fig:E_t} suggests a linear fit of the time-dependent average energies as
\begin{equation}\label{eq:E_drift}
\displaystyle \left\langle \Ecal^n \right\rangle_{\De t} = E_0\left(1 + \al_{\De t} n \De t\right),
\end{equation}
which allows to identify the relative energy drift rate $\al_{\De t}$. 
Figure \ref{fig:dE_dt} presents the dependence of $\al_{\De t}$ as a function of the timestep.

The SER scheme has energy drift rates of the same order than SSA, whereas the drift rates for SVV are substantially larger in absolute value. In addition, we notice that the drift rates of SSA and SER schemes remain lower than $5\times10^{-6}$, except for SER at $\De t\geq0.007$. This means that the energy variation would be less than $5\%$ for simulation times lower than $t_{\rm max} = 10^4$. Such simulation times may be long enough to estimate equilibrium properties, but they may prove somewhat short to estimate transport properties.

Moreover, the results of Figure \ref{fig:dE_dt} suggest that the drift rate $\al_{\De t}$ has a polynomial behavior with respect to~$\De t$ :
\begin{displaymath}
\displaystyle \al_{\De t} = K\De t^\ka.
\end{displaymath}
A least-square fit in a log-log diagram gives $\ka \simeq 2$ for SVV, and $\ka \simeq 4$ for the other schemes. This fast increase of the drift rate places a severe limitation on the use of larger timesteps in DPDE simulations.

\subsubsection{Drifts of individual energy components}
\label{subsubsection:observable_drifts}

\begin{figure}[h!]
\center
\includegraphics[scale=\SCALEXGR]{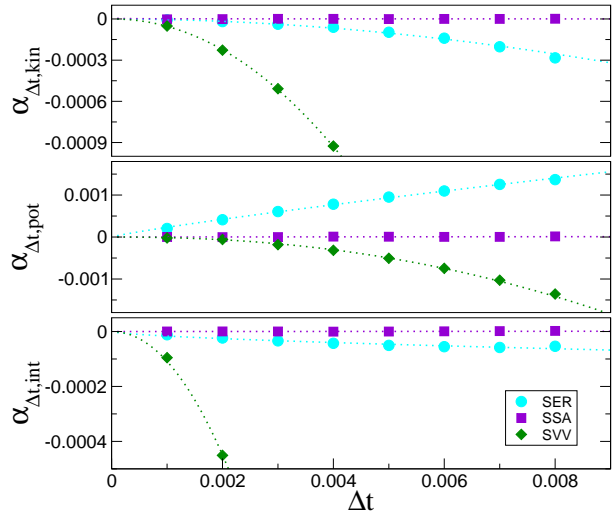}
\caption{\label{fig:dObs_dt} Drift rates per unit time of the kinetic (top), potential (middle) and internal (bottom) energies.}
\end{figure}

We studied in Section~\ref{subsubsection:energy_drifts} total energy drifts. Let us now have a more precise look at the drifts of individual energy components (kinetic, potential and internal), for which the drift rates as a function of the timestep are plotted in Figure~\ref{fig:dObs_dt}. As expected, the drift rates for SVV are large for all energy components, and small for all components for SSA. SER on the other hand exhibits a non-trivial behavior: its drift rates are quite large for individual components of the energy, although these large drifts compensate each other in order for the total energy to drift only slowly. In fact, we even observed in some situations that the drift of some energy components was not linear as a function of time. The very notion of drift rate for SER is therefore dubious (see Ref.~\onlinecite{PhDHomman} for further precisions on this specific issue).

\subsection{Projected dynamics to ensure stability}
\label{subsection:drift_correction}

In order to remove the energy drift, a natural approach, already suggested in Ref.~\onlinecite{LisalBrennan11}, is to enforce the energy conservation by an appropriate projection on a surface of constant energy. There are several possible projection procedures. 

One possible option is to resort to some Lagrange multiplier. A configuration $x=(q,p,\ve)$ such that $\Ecal(x)\neq E_0$ is replaced by $x - \la \na \Ecal(x)$, with $\lambda$ chosen such that
\begin{displaymath}
    \Ecal\Big(q-\la\na_q\Ecal(x),p-\la\na_p\Ecal(x),\ve-\la\na_{\ve}\Ecal(x)\Big) = E_0, 
\end{displaymath}
or more explicitly 
\begin{equation}\label{eq:projection_general}
    \Ecal\left(q-\la\nabla U(q),p-\frac{\la}m p,\ve-\la\vec1\right) = E_0,
\end{equation}
where $\vec1$ is the $N$-dimensional vector whose components are all equal to~1. However, the numerical computation of the parameter $\lambda$ satisfying~\eqref{eq:projection_general} is a computationally expensive task typically requiring several iterations of a Newton procedure, and hence several evaluations of the energy per timestep. Note also that the total momentum is not preserved, so that this extra conservation law should be subsequently enforced in an appropriate manner.

It is therefore much more convenient from a practical viewpoint to play on the internal energies only to adjust the total energy -- as was also already suggested in Ref.~\onlinecite{LisalBrennan11}. More precisely, at the end of one iteration of the numerical scheme, the resulting internal energies are replaced by $\ve^{n+1}_i - \big(\Ecal(\tde x^n)-E_0\big)/N$, where $\tde x^n$ is the unconstrained update of the configuration $x^n$. Note that this allows to remove energy variations due to the discretization of both the Hamiltonian and fluctuation/dissipation parts. In practice, one has to be careful that internal energies should however remain positive.

Enforcing the total energy conservation allows us to obtain a well-defined steady state, for which average properties can be safely computed. For the equilibrium simulations presented in the remainder of Section~\ref{section:equilibrium_simulations}, we use the SEM, SSA, SER and Hybrid schemes presented in Section~\ref{section:numerical_schemes}, but corrected by the projection procedure on the internal energies. 

\subsection{Timestep bias on average properties}
\label{subsection:projection}

\subsubsection{Sequential simulations}
\label{subsubsection:sequential_simulations}

\begin{figure}[h!]
\center
\includegraphics[scale=\SCALEXGR]{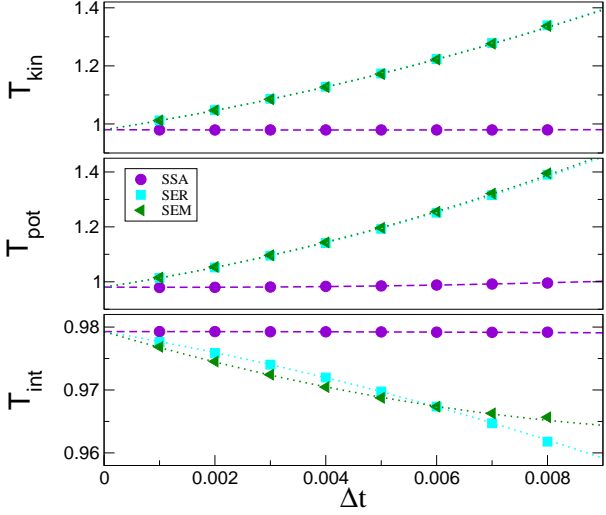}
\caption{\label{fig:obs_dt_proj_big} Numerical estimations of the equilibrium temperature as a function of the timestep for $\sig=4$.
Top: Kinetic temperature~\eqref{eq:Tkin} and potential temperature~\eqref{eq:Tcon}.
Bottom: Internal temperature~\eqref{eq:Tint}.}
\end{figure}

\begin{figure}[h!]
\center
\includegraphics[scale=\SCALEXGR]{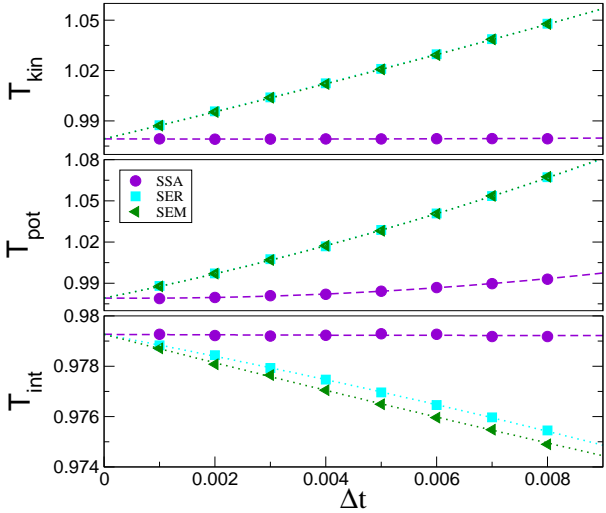}
\caption{\label{fig:obs_dt_proj_small} Numerical estimations of the equilibrium temperature as a function of the timestep for $\sig=2$.
Top: Kinetic temperature~\eqref{eq:Tkin} and potential temperature~\eqref{eq:Tcon}.
Bottom: Internal temperature~\eqref{eq:Tint}.}
\end{figure}

The results presented in this section are obtained by computing averages over one long trajectory of physical time $t_f=1000$. Figure~\ref{fig:obs_dt_proj_big} shows the average temperatures obtained from the estimators~\eqref{eq:Tkin}, \eqref{eq:Tcon} and~\eqref{eq:Tint} for $\sig=4$, while results for the smaller fluctuation strength $\sig = 2$ are reported in Figure~\ref{fig:obs_dt_proj_small}. 

As expected, all the estimations of the temperature extrapolate to the same value as $\De t\to 0$, for all schemes and for all components of the energy. The second point is that the systematic biases depend on the timestep $\De t$ as expected, but also on the value of~$\sig$: larger values of $\sigma$ lead to much larger errors.
The third point is that, in opposition to the biases of SSA which remain very small even for large values of $\De t$ and $\sig$, those of SEM and SER are much larger e.g, the bias on the estimation of the kinetic temperature for SER and SEM at $\De t=0.008$ and $\sig=4$ is around 30\% of the extrapolated reference value.
Moreover, in the case of large $\De t$ and large $\sig$ (typically $\sig=4$ and $\De t\geq 0.006$), SER and SEM biases are larger than the linear increase in~$\De t$ observed for smaller timesteps. In addition, for internal temperatures estimations, the biases between SER and SEM deviate at larger timesteps. 

\subsubsection{Parallel simulations}
\label{subsubsection:parallel_simulations}

\begin{figure}[h!]
\center
\includegraphics[scale=\SCALEXGR]{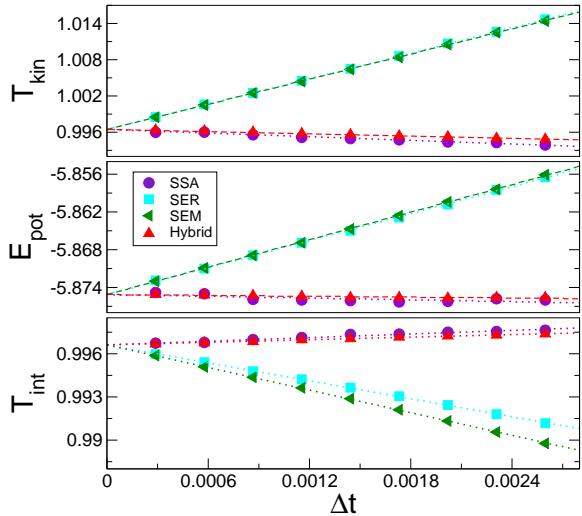}
\caption{\label{fig:obs_dt_proj_parallel} Numerical estimations of the equilibrium properties as a function of the timestep for $\sig=2$ obtained with parallel simulations.
Top: Kinetic temperature estimates~\eqref{eq:Tkin}.
Middle: Average potential energy $\langle U(q)\rangle$. 
Bottom: Internal temperature estimates~\eqref{eq:Tint}.
SSA estimates have been computed on a system of $N=4000$ particles on a single processor.}
\end{figure}

Thermodynamics averages in this section are computed over a single trajectory of total time $t_f=450$. The system was decomposed on $N_{\rm proc}=8$ processors. Figure~\ref{fig:obs_dt_proj_parallel} presents the temperature estimations as a function of the timestep for $\sigma=2$, compared to reference values computed by a sequential SSA simulation of a system of $N=4000$ particles.

Simulation results for $\sig=4$ are quite similar to sequential results (although with larger biases), except for SER.
Indeed, the scheme is not stable even for small timesteps since the energy reinjection procedure may lead to negative internal energies.
This is related to the lower value of $C_v$ used for parallel simulations.

CPU timings are reported in the second column of Table~\ref{table:benchmarks}. For the comparison to be fair, all the timings have been measured on a single core. We also compare these timings to reference timings from a sequential SSA simulation (third column of Table~\ref{table:benchmarks}).
The last column of Table~\ref{table:benchmarks} gives the required CPU time of a simulation of the same system as in Figure~\ref{fig:obs_dt_proj_parallel}, using a timestep chosen in order to obtain a given accuracy on the estimation of the kinetic temperature.
For each scheme, this timestep, denoted $\De t_{\rm lim}$, is taken such that the bias of the kinetic temperature estimation is equal to $e_{\rm lim}$, where $e_{\rm lim}$ is defined as the largest bias of the Hybrid kinetic temperature estimations of Figure~\ref{fig:obs_dt_proj_parallel}.
The biases are computed as $e_{\De t}=\left|\langle T_{\rm kin}\rangle_{\De t} - \langle T_{\rm kin}\rangle_{\De t=0}\right|$, where $\langle T_{\rm kin}\rangle_{\De t=0}$ is obtained by extrapolating the results of Figure~\ref{fig:obs_dt_proj_parallel} to $\De t=0$.
The timings of the last column of Table~\ref{table:benchmarks}, denoted $\tau^{T_{\rm kin}}_{\rm cpu}$, are then obtained by multiplying the timings of the second column of Table~\ref{table:benchmarks} with the number of iterations necessary to reach $t_f=100$ when using the timesteps $\De t_{\rm lim}$.
The choice of the kinetic temperature is arbitrary (as is the choice of $e_{\rm lim}$).
However, timings related to other observables show qualitatively similar results. The conclusion is that the Hybrid scheme allows to achieve an accuracy comparable to SSA for a given CPU cost, while being easily parallelizable. On the other hand, the SER scheme on its own, or SEM, are not sufficiently accurate to provide interesting alternatives.

\begin{table}[h!]
    \begin{tabular}{|c|c|c|l|}
    \toprule
    scheme                  & $t_{\trm{cpu}}$                   & ratio to SSA    & $\tau^{T_{\rm kin}}_{\rm cpu}$ \\
    \midrule
    SSA                     & $3.72\times\num{e-5}$~$\si{\second}$ & \tbf{1   }   & $ 1.82$~$\si{\second}$\\
    SEM                     & $3.35\times\num{e-5}$~$\si{\second}$ & \tbf{0.90}   & $12.09$~$\si{\second}$\\
    SER                     & $7.85\times\num{e-5}$~$\si{\second}$ & \tbf{2.11}   & $28.34$~$\si{\second}$\\
    Hybrid                  & $6.40\times\num{e-5}$~$\si{\second}$ & \tbf{1.72}   & $ 2.13$~$\si{\second}$\\
    \bottomrule
    \end{tabular} 
    \caption{\label{table:benchmarks} Column 2: average CPU time per particle per timestep and per processor. Column 3: timings of column 2 divided by the SSA timing. Column 4: Required CPU time corresponding to simulations with timesteps such that the kinetic temperature estimation has a given bias. This bias is chosen to be the largest bias of the Hybrid kinetic temperature estimations of Figure~\ref{fig:obs_dt_proj_parallel}.}
\end{table}

We see from Figure~\ref{fig:obs_dt_proj_parallel} that, as in Section~\ref{subsubsection:sequential_simulations}, SEM and SER estimations are very similar.
However, Table~\ref{table:benchmarks} tells us that SER is much slower than SEM, and that using SER takes more than twice as long as using SEM in order to reach a given accuracy.
This subperformance of SER is tempered by the fact that we correct here all schemes with the projection of Section~\ref{subsection:drift_correction}. For the Hybrid scheme, Figure~\ref{fig:obs_dt_proj_parallel} shows that it yields much smaller biases on average properties, comparable to those of SSA.
This increased accuracy more than compensates the extra computational cost compared to a very simple scheme such as SEM, as it can be seen in Table~\ref{table:benchmarks}.

\subsection{Influence of the parallelization on the Hybrid scheme}
\label{subsection:parallelization_influence}

There are no other parameters than the timestep for SSA, SER and SEM schemes. This is not the case for the Hybrid scheme, where the spatial redistribution of the simulation box between processors affects the computation.
It is thus necessary, in order to further validate the Hybrid scheme, to study the influence of the ratio $N/N_{\rm proc}$, where $N_{\rm proc}$ is the number of processors used for the simulation.
Indeed, the ratio $N/N_{\rm proc}$ directly determines the fraction of interactions treated with SER or SSA.

\begin{figure}[h!]
    \center
    \includegraphics[scale=0.60]{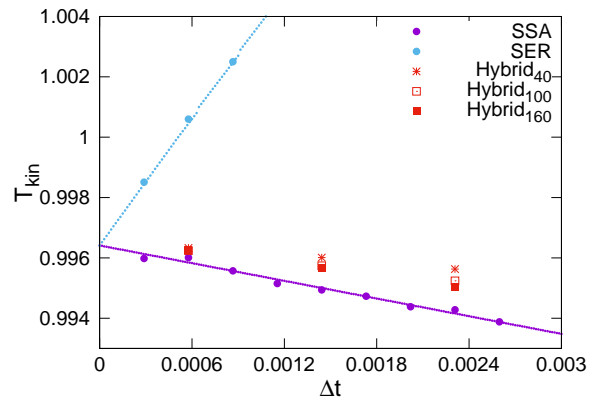}
    \caption{\label{fig:Nproc_ratio_influence} Kinetic temperature estimations for Hybrid simulations with a number $N_x$ of particles along the $x$ axis equal to $40$, $100$ and $160$. The SSA and SER estimations displayed are those reported in Figure~\ref{fig:obs_dt_proj_parallel}.}
\end{figure}

In order to study this influence, Hybrid simulations are performed on $N_{\rm proc}=8$ processors for systems in the same thermodynamic state as those of Section~\ref{subsubsection:parallel_simulations}.
The timesteps of the simulations range from $\De t=5\times\num{e-4}$ to $\De t=2.5\times\num{e-3}$ and the number of particle from $40\times10\times10$ to $160\times10\times10$.
Note that the parallelization is performed in the $x$ direction, and we only change the number of particles along this axis.
Averages obtained with the Hybrid scheme are presented in Figure~\ref{fig:Nproc_ratio_influence}, together with the reference SSA and SER estimations of Figure~\ref{fig:obs_dt_proj_parallel}.
The $N/N_{\rm proc}$ ratios according to the system sizes are displayed in Table~\ref{table:NNproc_ratio}.

\begin{table}[h!]
    \begin{tabular}{|c|c|l|}
    \toprule
    system size           & $N/N_{\rm proc}$ \\
    \midrule
    $ 40\times10\times10$ & $ 500$           \\
    $100\times10\times10$ & $1250$           \\
    $160\times10\times10$ & $2000$           \\
    \bottomrule
    \end{tabular} 
    \caption{\label{table:NNproc_ratio} $N/N_{\rm proc}$ ratio according to the system size.}
\end{table}

We notice in Figure~\ref{fig:Nproc_ratio_influence} that, as expected, Hybrid estimations go from those of SER towards those of SSA as $N/N_{\rm proc}$ increases.
In the particular situation considered here, SER and SSA estimations have biases of opposite signs that compensate in the Hybrid scheme, thus leading to biases smaller than those of SSA.
Averages of the internal temperature exhibit similar behaviors as in Figure~\ref{fig:Nproc_ratio_influence}.
SSA and Hybrid biases on the potential energy are almost null whatever the ratio $N/N_{\rm proc}$.

%------------------------------------------------------------------------
\section{Accuracy on dynamical properties: the example of shock waves}
\label{section:shock_simulations}
  
We study in this section the dynamical properties of systems out of equilibrium, using the schemes presented in Section~\ref{section:numerical_schemes}. The system we consider is composed of $N=450,000$ DPDE particles. In order for the numerical values to be easier to interpret from a physical viewpoint, we work in this section in physical units. The Lennard-Jones parameters in~\eqref{eq:Ep} are those of Argon: $\ve_{\rm LJ}=1.657\num{e-21}$~J and $r_{\rm LJ}=3.40510^{-10}$~m, while the masses of the particles are set to $m=6.634\times\num{e-26}$~kg. The initial condition is obtained by melting a face centered cubic crystal composed of $250\times15\times15=112,500$ unit cells, at density $\rho = 1228$~kg.m$^{-3}$, with periodic boundary conditions. We use $50\times2\times2=200$ cores for the simulations, and various timesteps ranging from $\De t=\num{e-15}$~s to $\De t=4\times \num{e-15}$~s (the value $\De t=\num{e-15}$~s corresponds to $\De t^* = 4.5\num{e-4}$ in reduced units). A timestep of $\De t=\num{e-15}$ is similar to the one used to integrate Hamiltonian dynamics of shock waves in Argon, but is already one order of magnitude larger than for full atom simulations of molecular systems. Finally, we set $C_v=1.381\times10^{-22}$~J.K$^{-1}$.

An important point in this section is that we no longer correct the schemes by the projection of Section~\ref{subsection:drift_correction} since we consider a nonequilibrium system which is inhomogeneous in space. This makes indeed the global energy reinjection procedure dubious.

We equilibrate the system at a given temperature $T_{\rm ref}=1000$~K by superimposing to the DPDE equations a Langevin thermostat on the momenta. With the notation introduced in~\eqref{eq:alternative_FDR}, we set the friction to $\ga=10^{-13}$~kg.s$^{-1}$ (which corresponds to $\sigma=5.255\times10^{-17}$~kg.m.s$^{-3/2}$ or $\sigma=7.37$ in reduced units).
Once equilibration is performed, we remove the periodic boundary conditions in the $x$ direction, and put two walls of fixed Lennard-Jones particles of infinite masses on the sides of the simulation box to confine the system. We next set the system in motion towards the left wall by adding a velocity equal to $u_p=-2000$~m.s$^{-1}$ to all the particles and to the right wall. The velocity of the right wall is maintained at $u_p$ throughout the simulation. This leads to the apparition of a shock wave propagating from the left to the right of the simulation box, with an average null velocity in the shocked state.

\begin{figure}[h!]
\center
\includegraphics[scale=0.6]{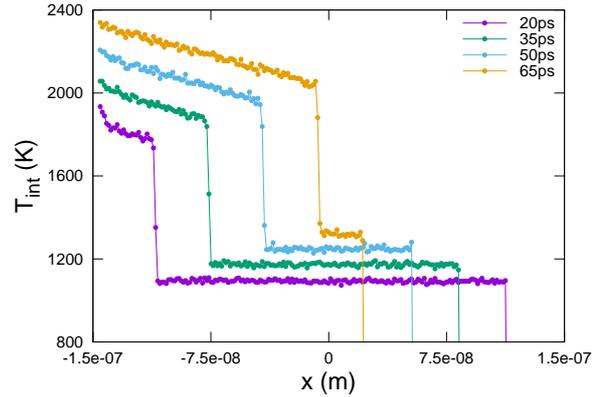}
\caption{\label{fig:shock_SEM} Instantaneous internal temperature profiles on the slabs of the cross-section $(y,z)$ for the SEM scheme integrated with a timestep $\De t=\num{e-15}$~s.}
\end{figure}

Figure~\ref{fig:shock_SEM} displays simulation results obtained with the SEM scheme with a small timestep $\De t=\num{e-15}$~s. Note that no stationary profile is obtained as the internal temperature increases in time. A similar increase is observed for the kinetic temperature. This increase is an artifact of the numerical scheme: we indeed confirmed by additional simulations (not reported here) that the increase in the temperature is more pronounced for larger timesteps.

\begin{figure}[h!]
\center
\includegraphics[scale=0.6]{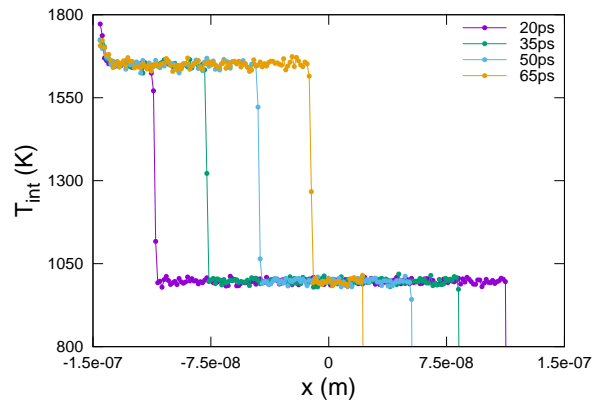}
\caption{\label{fig:shock_new} Instantaneous internal temperature profiles on the slabs of the cross-section $(y,z)$ for the Hybrid scheme integrated with a timestep $\De t=\num{e-15}$~s.}
\end{figure}

On the other hand, for shock simulations performed with the SER or Hybrid scheme, no such drift is observed. Figure~\ref{fig:shock_new} presents Hybrid internal temperature profiles for various times, for a simulation performed with the same timestep $\De t=\num{e-15}$~s. Let us also mention that results obtained with SER (not reported here) are completely comparable to the ones presented in Figure~\ref{fig:shock_new}. 

\begin{figure}[h!]
\center
\includegraphics[scale=0.6]{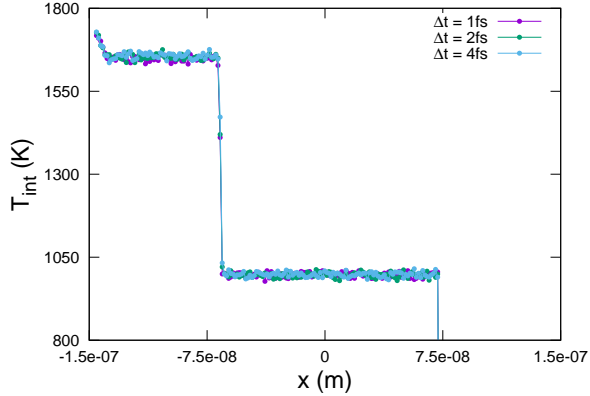}
\caption{\label{fig:shock_Hybrid_alldt} Instantaneous internal temperature profiles on the slabs of the cross-section $(y,z)$ for the Hybrid scheme integrated with timesteps ranging from $\De t=\num{e-15}$~s to $\De t=4 \times \num{e-15}$~s at time $t=40$~$\si{\pico\second}$.}
\end{figure}

Finally, the results are very similar for Hybrid simulations with timesteps up to $\De t=4\times\num{e-15}\si{\second}$, as reported in Figure~\ref{fig:shock_Hybrid_alldt}.

This shows that the schemes we developed for equilibrium simulations, in particular the Hybrid scheme, can be used without any projection procedure to simulate dynamical evolutions out of equilibrium, with timesteps which are even larger than the ones used for standard deterministic dynamics.

%------------------------------------------------------------------------
\section{Conclusion}

We presented two new numerical schemes for the integration of DPDE, based on a splitting strategy between the Hamiltonian part of the dynamics and the fluctuation/dissipation. The first scheme, SER, is constructed by decomposing the kinetic energy variations into elementary contributions related to each particle. This allows to adjust the internal energies in order for the total energy to be constant. The decomposition into elementary contributions can be seen as a discrete counterpart to the It\^o calculus used for the derivation of the dynamics in the internal energies. The other scheme, namely Hybrid, is a blending of SSA and SER scheme, where a SER discretization is applied to particle couples belonging to different processors.

We carefully compared the ability of these new schemes to compute equilibrium properties. All schemes exhibit some spurious energy drift over time, which can however be straightforwardly corrected by an appropriate projection (performed here by adjusting the internal energies\cite{LisalBrennan11}).
The scheme obtained by applying a projection to Hybrid has an accuracy quite comparable to the reference accuracy provided by SSA.
Therefore, Hybrid is a realistic option for a massively parallel DPDE simulation -- in contrast to other straightforwardly parallelizable schemes where the fluctuation/dissipation is integrated with a simple Euler discretization.
The SER was found to be less satisfactory since the individual components of the energy show quite large biases.
In addition, SER was found to be less stable than SEM or SSA when corrected schemes are considered.
However, its blending into the Hybrid scheme cures this somewhat degraded accuracy and stability.
Moreover, SER, in opposition to SSA or Hybrid, is threadable, thus compatible with the future architectures of supercomputers.

We finally validated the dynamical behavior of our schemes on a non-equilibrium simulation of a shocked Lennard-Jones material performed on hundreds of processors, a simulation impossible with a naive discretization of DPDE. In our opinion, this represents the strongest contribution of this work since it shows that the schemes we introduced, in particular the Hybrid scheme, allow to easily simulate dynamical evolutions of large scale systems on massively parallel architectures while at the same time being very accurate on the prediction of equilibrium properties.

%------------------------------------------------------------------------
\section*{Acknowledgments}

This work is supported by the European Research Council under the European Union's Seventh Framework Program (FP/2007-2013) / ERC Grant Agreement number 614492.

%-----------------------------------------------------------------------
\appendix

\section{Stochastic Velocity-Verlet}
\label{sec:SVV}

The Stochastic Velocity-Verlet algorithm, denoted SVV, is an adaption to the DPDE setting of the well-known Velocity-Verlet algorithm to integrate Hamiltonian dynamics\cite{Verlet67}. It consists in using a Strang splitting of DPDE, where the positions $q_i$ are updated once with a timestep $\De t$, and the momenta $p_i$ and internal energies $\ve_i$ are updated twice, both before and after the position update, with a timestep $\frac{\De t}2$:
\[
\left\{\begin{aligned}
p_i^{n+1/2} & = \pin- \na_{q_i}U(q^n)\frac{\De t}2 \\
& \ + \sum_{\substack{j=1\\j\neq i}}^N -\gaijn\chi(\rijn)\vijn\frac{\De t}2 + \sig\sqrt{\chi(\rijn)}\Gijn\frac{\sqrt{\De t}}2, \\
\ve_i^{n+1/2} & = \vein             + \sum_{\substack{j=1\\j\neq i}}^N \left(\gaijn(\vijn)^2 - d\frac{\sig^2}{m}\right)\chi(\rijn)\frac{\De t}2 \\
&  \                  - \sum_{\substack{j=1\\j\neq i}}^N \sig\vijn\sqrt{\chi(\rijn)}\Gijn\frac{\sqrt{\De t}}2,  \\
q_i^{n+1}       & = \qin              + \De t\frac{p_i^{n+1/2}}m, \\
p_i^{n+1}   & = p_i^{n+1/2}   - \na_{q_i}U(q^{n+1})\frac{\De t}2, \\
& \                   - \sum_{\substack{j=1\\j\neq i}}^N \gamma_{ij}^{n+1/2}\chi(r^{n+1}_{ij})v^{n+1/2}_{ij}\frac{\De t}2 \\
& \                  + \sum_{\substack{j=1\\j\neq i}}^N \sig\sqrt{\chi(r^{n+1}_{ij})}\Gijn\frac{\sqrt{\De t}}2, \\
\ve_i^{n+1}       & =  \ve_i^{n+1/2} \\
& \ + \sum_{\substack{j=1\\j\neq i}}^N \left(\gamma_{ij}^{n+1/2} \left(v^{n+1/2}_{ij}\right)^2 - d\frac{\sig^2}{m}\right)\chi(r^{n+1}_{ij})\frac{\De t}2 \\
& \                   - \sum_{\substack{j=1\\j\neq i}}^N \sig v^{n+1/2}_{ij}\sqrt{\chi(r^{n+1}_{ij})}\Gijn\frac{\sqrt{\De t}}2.
\end{aligned}\right.
\]
Note that the same random number is used in the first and second updates of the momenta. However, the fluctuation/dissipation part has to be computed twice per step. 

\section{Consistency of numerical schemes}
\label{sec:scheme_consistency}

There are several notions of consistency for Stochastic Differential Equations (SDEs), see for instance Ref.~\onlinecite{BookMilsteinTretyakov}. In this work, we consider the consistency of numerical schemes in the sense of weak errors. As a corollary, this allows to estimate errors on average properties when the schemes are ergodic.

In order to be more precise, we define the evolution operator associated with the SDE~\eqref{eq:DPDE}:
\begin{equation}\label{eq:SDE_flow_map}
    \displaystyle \left(\mrm{e}^{t\Lcal}A\right)(x) = \Ebb\left[ A(x_t) | x_0=x \right],
\end{equation}
where $x=(q,p,\ve)$ and $\Ebb$ denotes the average over all realizations of the Brownian motions in~\eqref{eq:DPDE}. In words, the evolution operators give the average value at time~$t$ of an observable~$A$, for a system started in a given configuration~$x_0$ at time~0. The operator $\Lcal$ is called the infinitesimal generator, and is the adjoint of the Fokker-Planck operator associated with the dynamics. For DPDE, it writes\cite{Espanol97}
\begin{equation}\label{eq:DPDE_generator}
\begin{aligned} \Lcal & = \sum\limits^N_{i=1} \frac{p_i}m\cdot\na_{q_i} - \na_{q_i}U\cdot\na_{p_i} \\
& \ \ + \frac12\sum\limits^N_{\substack{j=1\\j\neq i}} \chi(\rij)\mathcal{A}_{ij}\left( -\gaij\vij + \frac{\sigij^2}2\mathcal{A}_{ij} \right), 
\end{aligned}
\end{equation}
where
\begin{equation}\label{eq:operator_Acalij}
\displaystyle \mathcal{A}_{ij} = \left(\na_{p_i} - \na_{p_j}\right) + \frac{\vij}2\left(\partial_{\ve_i}+\partial_{\ve_j}\right).
\end{equation}

The numerical schemes we consider in this work can be represented by a mapping $\Phi_{\De t}(x,G)$ as $x^{n+1} = \Phi_{\De t}(x^n,G^n)$. The evolution operator $P_{\De t}$ associated to a numerical scheme is defined as
\begin{equation}\label{eq:evolution_operator}
\displaystyle \left(P_{\De t}A\right)(x) = \Ebb\left[\left.A(x^{n+1})\right|x^n=x \right],
\end{equation}
where $\Ebb$ now is an average over all the random variables involved in the computation of $x^{n+1}$.

The weak error of a numerical scheme is the maximum error between the time averages of the exact solution and the ones predicted with the numerical scheme. A method is of weak order~$\eta$ if, for any observable $A$ and for any simulation time $T=\Niter\De t$, there exists $K>0$ such that, for $\De t$ sufficiently small,
\begin{equation}\label{eq:weak_error}
\sup_{0\leq n\leq \Niter} \Big| \Ebb\left[A(x_{n\De t})\right] - \Ebb\left[A(x^n)\right] \Big| \leq K\De t^\eta,
\end{equation}
where averages are taken for a given initial condition~$x_0 = x^0$. In order to be of weak error~$\eta$, a numerical scheme should satisfy\cite{BookMilsteinTretyakov}
\begin{equation}\label{eq:weak_consistency}
    \displaystyle \left(P_{\De t}A\right)(x) = \left(\mrm{e}^{\Lcal\De t}A\right)(x) + \Ocal(\De t^{\eta+1}).
\end{equation}
In fact, when this condition holds, it can be proved that, provided the numerical scheme and the continuous dynamics are ergodic, the invariant measure of the numerical scheme is correct up to errors at most $\mathrm{O}(\De t^\eta)$ (and sometimes less, see for instance Ref.~\onlinecite{LMS15} for a discussion in the context of Langevin dynamics). Therefore, the exponent~$\eta$ in~\eqref{eq:obs_estimation_theorique} is determined by expansions such as~\eqref{eq:weak_consistency}.

We prove below that~\eqref{eq:weak_consistency} holds with $\eta=1$ for SER (see Section~\ref{sec:consistency_SER}). This result is obtained by a general approximation result for splitting schemes, recalled in Section~\ref{sec:general_consistency}.

\subsection{A general consistency result}
\label{sec:general_consistency}

We recall in this section the general strategy for proving~\eqref{eq:weak_consistency} with $\eta=1$ for splitting schemes. We assume that the dynamics can be decomposed into $m$ elementary dynamics. This induces a decomposition of the generator as
\begin{displaymath}
\mcal L = \Lcal_1 + \hdots + \Lcal_m.
\end{displaymath}
Let $P_{k,\De t}$ be the evolution operator associated with the numerical integration of the elementary subdynamics with generators $\Lcal_k$. If the splitting is done according to the Trotter formula\cite{Trotter59}
\begin{equation}\label{eq:Trotter_formula}
P_{\De t} = P_{1,\De t} \hdots P_{m,\De t},
\end{equation}
and if~\eqref{eq:weak_consistency} holds for every subdynamics with $\eta=1$, then~\eqref{eq:weak_consistency} holds for the complete numerical scheme~$P_{\De t}$ (see for instance the discussion in Ref.~\onlinecite{LMS15}). 

The Hamiltonian part of DPDE is integrated with a Velocity-Verlet algorithm\cite{Verlet67}, and the corresponding evolution operator turns out to be correct at second order. Therefore, in order to prove the weak consistency of order~1 for the schemes under consideration, it suffices to prove that the fluctuation/dissipation is integrated with a scheme of weak order~1.

\subsection{Consistency of SER}
\label{sec:consistency_SER}

A simple computation shows that~\eqref{eq:SER_fd} implies
\begin{equation}\label{eq:SER_ve}
\displaystyle \ve^{n+1}_i = \vein - \frac 12  \sum^N_{\substack{j=1\\j\neq i}} \de \ve_{ij}^n + \Ocal(\De t^2),
\end{equation}
where
\[
\begin{aligned}
\de \ve_{ij}^n & = -\gaijn(v^n_{ij})^2\chi(\rijn)\De t + \sig v^n_{ij}\sqrt{\chi(\rijn)}\Gijn\sqrt{\De t} \\
& \qquad + B^n_{ij}\De t + C^n_{ij}\De t^{3/2}.
\end{aligned}
\]
In this expression, we introduced
\begin{displaymath}
\left\{\begin{aligned}
    B^n_{ij} &=\displaystyle \frac1{2m}R^n_{ij}(R^n_i-R^n_j), \\
    C^n_{ij} &=\displaystyle -\frac1{2m}\left( F^n_{ij}\left(R^n_i-R^n_j\right) + R^n_{ij}\left(F^n_i-F^n_j\right) \right),
\end{aligned}\right.
\end{displaymath}
with
\begin{displaymath}
\left\{\begin{array}{llll}
F^n_{ij}     &=\displaystyle \ga^n_{ij}v^n_{ij}\chi(r^n_{ij}),        & F^n_i &=\displaystyle \sum^N_{\substack{j=1\\j\neq i}} F^n_{ij}, \\
R^n_{ij}     &=\displaystyle \sig \sqrt{\chi(r^n_{ij})} G^n_{ij},     & R^n_i &=\displaystyle \sum^N_{\substack{j=1\\j\neq i}} R^n_{ij}.
\end{array}\right.
\end{displaymath}
The random variables $C^n_{ij}$ involve only terms containing one Gaussian random variable. Therefore, they vanish in average: $\Ebb\left[ C^n_{ij} \right] = 0$. The random variables $B^n_{ij}$ involve terms containing products of Gaussian variables, but their averages can be easily computed as $\Ebb[G^n_{ij}G^n_{kl}] = d \, \de_{ik}\de_{jl}$. A simple computation then shows that $\Ebb\left[ B^n_{ij} \right] = d\sig^2\chi(\rijn)/m$.

Let $\Phi^{\trm{EE}}_{\De t}(x,G) = \left(\Phi^{\trm{EE},p}_{\De t}(p,G),\Phi^{\trm{EE},\ve}_{\De t}(\ve,G)\right)$ be the result of an Euler-Maruyama discretization of the fluctuation/dissipation dynamics~\eqref{eq:DPDE_dissipative} (\emph{i.e.} the scheme~\eqref{eq:SEM_fd}). The previous computations show that SER can be seen as a perturbation of the Euler-Maruyama scheme: $p^{n+1}_i   = \Phi^{\trm{EE},  p}_{\De t}(  p^n,G^n)_i,$ and 
\[
\begin{aligned}
& \ve^{n+1}_i = \Phi^{\trm{EE},\ve}_{\De t}(\ve^n,G^n)_i \\
& - \frac12\sum_{\substack{j=1\\j\neq i}} \left( B^n_{ij}-\frac{d\sig^2}m\chi(\rijn) \right)\De t + C^n_{ij}\De t^{3/2} + \Ocal(\De t^2).
\end{aligned}
\]
Since the Euler-Maruyama scheme is weakly consistent of order~1, and since the two dominant terms in the above expansion vanish in average, a simple Taylor expansion then shows that the integration of the fluctuation/dissipation part in SER is also  weakly consistent of order~1.

%------------------------------------------------------------------------


\begin{thebibliography}{10}

\bibitem{HoogerbruggeKoelman92}
P.~J. Hoogerbrugge and J.~M. V.~A. Koelman. 
\newblock {\em Europhys. Lett.}, 19(3):155, 1992.

\bibitem{EspanolWarren95}
P.~Espa{\~n}ol and P.~Warren.
\newblock {\em Europhys. Lett.}, 30(4):191, 1995.

\bibitem{MoeendarbaryNG09}
E.~Moeendarbary, T.~Y.~NG and M.~Zangeneh.
\newblock{\em Int. J. Appl. Mech.}, 1(4):737, 2009.

\bibitem{DischerOrtiz07}
D.~E.~Discher, V.~Ortiz, G.~Srinivas, M.~L.~Klein, Y.~Kim, D.~Christian, S.~Cai, P.~Photos and F.~Ahmed.
\newblock{\em Prog. Polym. Sci.}, 32:838, 2007.

\bibitem{GoujonMalfreyt11}
F.~Goujon, P.~Malfreyt and D.~J.~Tildesley.
\newblock{\em J. Chem. Phys.}, 129(3):034902, 2007.

\bibitem{AvalosMackie97}
J.~Bonet Avalos and A.~D. Mackie.
\newblock {\em Europhys. Lett.}, 40(2):141, 1997.

\bibitem{Espanol97}
P.~Espa{\~n}ol.
\newblock {\em Europhys. Lett.}, 40(6):631, 1997.

\bibitem{Stoltz06}
G.~Stoltz.
\newblock {\em Europhys. Lett.}, 77(8):849--855, 2007.

\bibitem{MSS07}
J.-B. Maillet, L. Soulard and G. Stoltz. 
\newblock {\em Europhys. Lett.}, 78(6):68001, 2007.

\bibitem{MVDS11}
J.-B. Maillet, G. Vallverdu, N. Desbiens and G. Stoltz. 
\newblock {\em Europhys. Lett.}, 96(6):68007, 2011. 

\bibitem{GrootWarren97}
R.~D. Groot and P.~B. Warren.
\newblock {\em J. Chem. Phys.}, 107(11):4423, 1997.

\bibitem{PagonabarragaHagen98}
I.~Pagonabarraga, M.~H.~J. Hagen, and D.~Frenkel.
\newblock {\em Europhys. Lett.}, 42(4):377, 1998.

\bibitem{BesoldVattulainen00}
G.~Besold, I.~Vattulainen, M.~Karttunen, and J.~M. Polson.
\newblock {\em Phys. Rev. E}, 62:7611, 2000.

\bibitem{VattulainenKarttunen02}
I.~Vattulainen, M.~Karttunen, G.~Besold, and J.~M. Polson.
\newblock {\em J. Chem. Phys.}, 116(10):3967, 2002.

\bibitem{LS15}
B. Leimkuhler and X. Shang.
\newblock {\em J. Comput. Phys.}, 280:72, 2015

\bibitem{LarentzosBrennan14}
J.~P. Larentzos, J.~K. Brennan, J.~D. Moore, M.~Lisal, and W.~D. Mattson.
\newblock {\em Comput. Phys. Commun.}, 185(7):1987, 2014.

\bibitem{AbuNada10}
A-N. Eiyad.
\newblock {\em Phys. Rev. E}, 81:056704, 2010.

\bibitem{AbuNada11}
A-N. Eiyad.
\newblock {\em Mol. Simulat.}, 37(2):135, 2011.

\bibitem{Shardlow03}
T.~Shardlow.
\newblock {\em SIAM J. Sci. Comput.}, 24(4):1267, 2003.

\bibitem{LisalBrennan11}
M.~Lisal, J.~K. Brennan, and J.~Bonet Avalos.
\newblock {\em J. Chem. Phys.}, 135(20):204105, 2011.

\bibitem{SDPD}
P. Espa{\~n}ol and M. Revenga.
\newblock {\em Phys. Rev. E}, 67:026705, 2003.

\bibitem{FMS15}
G. Faure, J.-B. Maillet and G. Stoltz.
\newblock in preparation.

\bibitem{SH04} 
A. Strachan and B.L. Holian. 
\newblock {\em Phys. Rev. Lett.} 94:014301, 2004. 

\bibitem{BookLelievreRousset}
T.~Leli{\`e}vre, M.~Rousset, and G.~Stoltz.
\emph{Free Energy Computations. A Mathematical Perspective.}
\newblock Imperial College Press, 2010.

\bibitem{SY06} 
T. Shardlow and Y.B. Yan.
\newblock Stoch. Dynam., 6(1):123, 2006. 

\bibitem{ButlerAyton98}
B.~D. Butler, G. Ayton, Owen~G. Jepps, and D.~J. Evans.
\newblock {\em J. Chem. Phys.}, 109(16):6519, 1998.

\bibitem{MBAN99}
A.~D. Mackie, J. Bonet Avalos, and V.~Navas.
\newblock {\em Phys. Chem. Chem. Phys.}, 1(9):2039, 1999.

\bibitem{SerranodeFabritiis06}
M.~Serrano, G.~De Fabritiis, P.~Espa{\~n}ol, and P.V. Coveney.
\newblock {\em Math. Comput. Simulat.}, 72(2-6):190, 2006.

\bibitem{Verlet67}
L.~Verlet.
\newblock {\em Phys. Rev.}, 159:98, 1967.

\bibitem{BruengerBrooks84}
A.~Bruenger, C.~L.~Brooks III, and M.~Karplus.
\newblock {\em Chem. Phys. Lett.}, 105(5):495, 1984.

\bibitem{AfsharSchmid12}
Y.~Afshar, F.~Schmid, A.~Pishevar, and S.~Worley.
\newblock {\em Comput. Phys. Commun.}, 184(4):1119, 2013.

\bibitem{PhDHomman}
A-A Homman.
\newblock PhD thesis, {\'E}cole Nationale des Ponts et Chauss{\'e}es, 2016.

\bibitem{BookMilsteinTretyakov}
G.~N. Milstein and M.~V. Tretyakov.
\newblock \emph{Stochastic Numerics for Mathematical Physics}, Springer, 2004.

\bibitem{LMS15}
B.~Leimkuhler, C.~Matthews, and G.~Stoltz.
\newblock {\em IMA J. Numer. Anal.}, in press, 2015.

\bibitem{Trotter59}
H.~F. Trotter.
\newblock {\em Proc. Am. Math. Soc.}, 10(4):545, 1959.

\end{thebibliography}
\end{document}